\documentclass{article}

\usepackage{graphics}
\usepackage{emulateapj}

\newbox\grsign \setbox\grsign=\hbox{$>$} \newdimen\grdimen \grdimen=\ht\grsign
\newbox\simlessbox \newbox\simgreatbox
\setbox\simgreatbox=\hbox{\raise.5ex\hbox{$>$}\llap
     {\lower.5ex\hbox{$\sim$}}}\ht1=\grdimen\dp1=0pt
\setbox\simlessbox=\hbox{\raise.5ex\hbox{$<$}\llap
     {\lower.5ex\hbox{$\sim$}}}\ht2=\grdimen\dp2=0pt
\def\simgreat{\mathrel{\copy\simgreatbox}}
\def\simless{\mathrel{\copy\simlessbox}}
 
\newcommand{\etal}{{et al.}\ }

\newcommand{\lbol}{L_{\mathrm X}}

\newcommand{\rCC}{r_{200} }
\newcommand{\MCC}{M_{200} }
\newcommand{\VCC}{V_{200} }
\newcommand{\Vcirc}{V_{\mathrm circ} }
\newcommand{\rhotilde}{\tilde{\rho}}
\newcommand{\rhocrit}{\rho_{\mathrm crit}}
\newcommand{\rhodm}{\rho_{\mathrm dm}}
\newcommand{\sigmaav}{\langle \sigma^2_{\mathrm dm} \rangle }
\newcommand{\sigmadm}{\sigma_{\mathrm dm}}
\newcommand{\sigmatilde}{\tilde{\sigma}}
\newcommand{\Tav}{\langle T \rangle}
\newcommand{\rhogas}{\rho_{\mathrm gas}}

\newcommand{\betajeans}{\beta_{\mathrm J}}

\newcommand{\cmin}{2.5}
\newcommand{\cmax}{9.5}
\newcommand{\nbest}{-1.1}

\begin{document}

\submitted{Submitted March 24, 1999; Accepted May 20, 1999 for
publication
in \emph{The Astrophysical Journal Letters}}
\title{The Cluster $L_X - \sigma$ Relation Has Implications for Scale-Free 
Cosmologies}
\author{Andisheh Mahdavi\altaffilmark{1}}
\altaffiltext{1}{amahdavi@cfa.harvard.edu}
\affil{Harvard-Smithsonian Center for Astrophysics, 60 Garden St.,
Cambridge, MA 02138}
\authoremail{amahdavi@cfa.harvard.edu}
 
\lefthead{Mahdavi}
\righthead{Cosmological $L_X - \sigma$ Relation}

\begin{abstract}

I show that the cluster $L_X - \sigma$ relation should be sensitive to
cosmologies with a scale-free power spectrum of initial density
fluctuations, $P(k) \propto k^n$. I derive the dependence, and argue
that a conservative interpretation of current observations implies $n
< -2.0$ and $n < \nbest$ at the one-sided 90\% and 99\% confidence
levels, respectively. This result, which agrees with constraints on
$n$ from the x-ray cluster temperature function, should be roughly
independent of the value of $\Omega$ or $\Lambda$.

\end{abstract}

\section{Introduction}

The dynamical state of a system of galaxies may be characterized by
three intimately related physical quantities: the bolometric x-ray
luminosity $\lbol$, the average emission-weighted plasma temperature
$\Tav$, and the average dark matter velocity dispersion, $\sigmadm$
(as traced by the projected velocity dispersion of the galaxies,
$\sigma_p$). There is an observed correlation among these parameters,
crudely given by $\Tav \propto \sigma_p^{\alpha_1}$, $\lbol \propto
\Tav^{\alpha_2}$, and $\lbol \propto \sigma_p^{\alpha_3}$, and roughly
in agreement with the predictions of physical models. My goal in this
letter is to show that the value of the slope $\alpha_{3}$ at $z=0$
can constrain the spectrum of the primordial density fluctuations,
$P(k) \propto k^n$.

One powerful tool for linking the x-ray properties of galaxy systems
with the cosmological parameters already exists. The cluster
temperature function, which is directly related to the cluster mass
distribution, can be used at zero redshift to measure $n$.  Henry \&
Arnaud (1991) analyze data from the Einstein satellite to obtain $n
\approx -1.7 \pm 0.55$; data from the ROSAT and ASCA missions
(Markevitch 1998), corrected for the presence of cooling flows,
implies a somewhat steeper $n \approx -2 \pm 0.3$. The redshift
evolution of the cluster temperature function can break the degeneracy
between the density parameter, $\Omega$, and the normalization of the
primordial spectrum (Henry 1997).

Here I introduce a new zero-redshift method. I suggest that
$\alpha_3$, the slope of the $L_X - \sigma_p$ relation, is related to
$n$ in a way that is complementary to the dependence of the
temperature function on $n$. While it is important to have a
flux-limited sample for the temperature function method, using
$\alpha_{3}$ to compute $n$ requires only a large number of pointed
observations of systems of galaxies. In \S \ref{sec:derive} I derive
the relationship between $\alpha_{3}$ and $n$. In \S \ref{sec:data} I
compare the results with available observations, and discuss possible
systematic biases. In \S \ref{sec:concl} I summarize.

\section{Derivation}
\label{sec:derive}

Here I use some of the definitions in the paper by Navarro, Frenk, \&
White (1997; NFW). My results, however, are largely independent of
their N-body simulations of dark matter halos, and apply to density
profiles different from the one they introduce.  Eke, Navarro, \&
Frenk (1998) provide a derivation of the $L_X - \Tav$ relation which
is analogous to part of what follows, but does not focus on any
cosmological use of $\alpha_2$.

\subsection{Halo Parameters and Definitions}

A spherically symmetric dark matter halo may be characterized by
$\rCC$, the radius which encloses 200 times the critical density of
the universe, $\rhocrit$. The characteristic mass, $\MCC$, is then
\begin{eqnarray}
\label{eq:m200}
\MCC & = & \frac{800 \pi}{3} \rCC^3 \rhocrit \\
\label{eq:rhocrit}
\rhocrit & = & \frac{3 H_0^2}{8 \pi G} Z(z), \\
Z(z,\Omega) & = & (1 + z)^3 \frac{\Omega_0}{\Omega(z)}.
\end{eqnarray}
Here $H_0$ is the Hubble constant, $\Omega$ is the density parameter,
and $z$ is the redshift. The halo's characteristic circular velocity
is $\VCC = \sqrt{G \MCC/\rCC}$, or, with the substitution of equation
(\ref{eq:m200}),
\begin{equation}
\label{eq:vcc}
\VCC = 10 H_0 \rCC Z^{1/2}.
\end{equation}
Now consider a halo density profile of the form,
\begin{equation}
\label{eq:rhodm}
\rhodm(r) = \rhocrit \delta_c \rhotilde(\frac{c r}{\rCC}), \\
\end{equation}
where $c$ is the halo concentration, $\delta_c$ is the characteristic
density, and $\rhotilde(y)$ is a nonnegative, declining function of
$y$. Then equation (\ref{eq:m200}) requires the following relation
between $\delta_c$ and $c$,
\begin{eqnarray}
\label{eq:cofc}
\delta_c & = & \frac{200 c^3}{3 C(c)}, \\
\label{eq:cofy}
C(y) & = & \int_0^y t^2 \rhotilde(t) dt.
\end{eqnarray}
In scale-free cosmologies, the characteristic density also depends on
$n$, the slope of the initial fluctuation spectrum, because $c$ should
trace the mean density of the universe at the time of the halo's
formation. NFW use Press-Schechter theory to examine
the exact relationship among $\delta_c$, $\MCC$, and $n$.  They find
that the characteristic density should be proportional to the mean
density of the universe at the epoch when the nonlinear mass $M_*$ is
(1) a fixed and (2) a very small fraction of the current halo
mass. This implies that $\delta_c$ should scale with $M$ and $n$ the
same way as the average background density scales with $M_*$ and $n$:
\begin{equation}
\label{eq:mofn}
\delta_c \propto \MCC^{-(n+3)/2}.
\end{equation}
The slope of this scaling relation is better than 5\% accurate for all
$\Omega\le 1$ cosmologies with either $\Lambda=0$ or
$\Omega+\Lambda=1$, where $\Lambda$ is the cosmological constant in
units of $3 H^2$. With the substitution $\gamma = -(n+3)/2$ and use of
equations (\ref{eq:m200}), (\ref{eq:vcc}), and (\ref{eq:cofc}),
\begin{equation}
\label{eq:vofn}
\frac{c^3}{C(c)} \propto \VCC^{3\gamma} Z^{-\gamma/2}.
\end{equation}

Next, I compute the circular velocity profile of the halo, $\Vcirc =
\sqrt{G M(r)/r}$. Under the transformation $r \rightarrow y r_{200} /
c$,
\begin{equation}
\Vcirc(y)^2 = \VCC^2 \frac{c C(y)}{y C(c)}
\end{equation}
Note that the dark matter velocity dispersion profile, $\sigmadm(y)$,
will not in general have the same shape as $\Vcirc(y)$, contrary to
the the assumption in similar derivations (e.g. Eke \etal
1998). Rather, the velocity dispersion of the halo is given by the
Jeans equation for a spherical, nonrotating system of collisionless
particles (Binney \& Tremaine 1987):
\begin{equation}
\frac{d \left( \sigma^2_r \rho \right) }{dr} + 2 \betajeans \rho
\sigma^2_r = - \frac{G M \rho}{r^2}.
\label{eq:jeans}
\end{equation}
Here $\sigma_r$ is the radial velocity dispersion, $\betajeans$ is
the velocity anisotropy parameter, and $M$ is the mass inside the
radius $r$. For systems with isotropic velocity dispersion tensors,
$\betajeans = 0$, and locally $\sigmadm^2 = 3
\sigma_r^2$. In terms of the previously defined quantities, equation
(\ref{eq:jeans}) has the solution
\begin{eqnarray}
\label{eq:sigofr}
\sigma_r^2(y) & =  & \VCC^2 \frac{c}{C(c)} \sigmatilde^2(y), \\
\sigmatilde^2(y) & = & \frac{1}{\rhotilde(y)}
\int_y^\infty \frac{\rhotilde(t) C(t)}{t^2} dt,
\end{eqnarray}
Finally, consider the average dark matter velocity dispersion,
$\sigmaav$, within the virial radius $\rCC$:
\begin{eqnarray}
\label{eq:sigav}
\sigmaav & = & \frac{\int_0^c 3 \sigma_r^2(y) \rhotilde(y) y^2 dy}
{\int_0^c \rhotilde(y) y^2 dy} \\
\label{eq:sigav2}
& = & \VCC^2 \frac{c D(c)}{C(c)^2}, \\
\label{eq:sigav3}
D(y) & = & \int_0^y 3 \sigmatilde^2(t) \rhotilde(t) t^2 dt.
\end{eqnarray}

\subsection{X-Ray Luminosity}

The x-ray luminosity of a ball of plasma is
\begin{equation}
\lbol = \int \lambda(T) n_e n_i dV.
\end{equation}
Here $\lambda(T)$ is the bolometric emissivity as a function of the
local temperature $T$, $n_e$ is the electron number density, $n_i$ is
the ion number density, and the integral is over the entire volume of
the system. Now if (1) $n_i = n_e$, (2) the system is spherically
symmetric, and (3) the gas density is related to the dark matter
distribution by $\rhogas(r) = f \rhodm(r)$, where $f$ is the gas mass
fraction, then
\begin{equation}
\lbol = 4\pi \int_0^\infty \lambda(T) \left( \frac{f \rhodm}{\mu m_p}
\right)^2 r^2 dr,
\end{equation}
where $\mu \equiv \rhogas / (n_i m_p)$ is the mean molecular
weight. Substitution of equation (\ref{eq:rhodm}) yields,
\begin{equation}
\lbol   =  4 \pi \left(\frac{f \rhocrit}{\mu m_p} \right)^2
\frac{\delta_c^2}{c^3} \rCC^3 \int_0^\infty \lambda(T) 
\rhotilde(y)^2 y^2 dy.
\end{equation}
Now eliminate $\rCC$, $\delta_c$, and $\rhocrit$ using equations
(\ref{eq:rhocrit}), (\ref{eq:vcc}) and (\ref{eq:cofc}):
\begin{eqnarray}
L_X & = & \frac{5 H_0 Z^{1/2}}{2 \pi G^2} 
\left(\frac{f}{\mu m_p} \right)^2  \frac{c^3}{C(c)^2} \VCC^3 \\
\nonumber
& \times & \int_0^\infty \lambda(T) \rhotilde(y)^2 y^2 dy.
\end{eqnarray}

Next, note that thermal bremsstrahlung is the dominant cooling process
for rich clusters of galaxies; hence $\lambda(T) \propto
T^{1/2}$. Then, if the gas is in local hydrostatic equilibrium with
the dark matter, $T(r) \propto \sigmadm(r)^2$, and hence $\lambda(T,r)
\propto \sigmadm(r)$. Leaving out the constants and substituting
equation (\ref{eq:sigofr}) yields,
\begin{equation}
\lbol \propto f^2 Z^{1/2} \frac{c^{7/2}}{C(c)^{5/2}} \VCC^4
\int_0^\infty 3 \sigmatilde(y) \rhotilde(y)^2 y^2 dy
\end{equation} 
The integral on the right-hand side is just a number, and may be
dropped.

Now one must replace $\VCC$, which is not observable, with the
projected galaxy velocity dispersion $\sigma_p$, which can be
determined from optical surveys. In the cores of relaxed clusters
$\sigma_p^2$ is proportional to $\sigmaav$, the average dark matter
velocity dispersion (equation \ref{eq:sigav}). Then
\begin{equation}
L_X \propto f^2 Z^{1/2} \frac{c^{3/2} C(c)^{3/2}}{D(c)^2} \sigma_p^4.
\end{equation}
If $f$ and $c$ remain constant, the above expression reduces to the
traditional $\lbol \propto \sigma^4$ scaling law from simpler,
dimensional arguments (e.g., Quintana \& Melnick 1982). However,
equation (\ref{eq:vofn}) tells us that $c$ is a function of the
velocity dispersion and the slope of the primordial spectrum. In fact,
once $\rhotilde(y)$ is specified, equations (\ref{eq:cofy}),
(\ref{eq:vofn}), (\ref{eq:sigav2}), and (\ref{eq:sigav3}) may be used
to eliminate $\VCC$, $c$, $C(c)$, and $D(c)$, and one may thus obtain
the dependence of $L_X$ on $\sigma_p$ and $n$. Specifically, suppose
that within the range of interest for $c$, $C(c)$ and $D(c)$ exhibit a
power law behavior of the form $C(c) \propto c^p$ and $D(c) \propto
c^q$. Then
\begin{equation}
\label{eq:final}
L_X \propto f^2 Z^{1/2 + \xi/6} \sigma_p^{4 - \xi},
\end{equation}
where
\begin{equation}
\label{eq:xi}
\xi = \frac{3 (3 + n) (3 + 3p - 4q)}{3 + 14 p - 9 q + n (6 p - 3 -
3q)}
\end{equation}

\subsection{Model Dependency}

I consider dark matter density profiles of the form $\rhotilde(y) =
y^{-a} (1 + y^b)^{-d}$. Thus each profile's shape may be specified by
a set of three numbers, $(a,b,d)$. Some common profiles and their
properties are listed in Table \ref{tbl:profiles}. 

Once the set $(a,b,d)$ is specified, $C(c)$ and $D(c)$ are readily
computable. The relevant range of $c$ for systems of galaxies comes
from measurements of surface number density profiles and of $\rCC$ in
clusters and groups of galaxies (Carlberg \etal 1997; Mahdavi \etal
1999). In these works, halos with masses in the range $10^{14} -
10^{16} M_\odot$ have concentrations $c \approx$ \cmin--\cmax, in good
agreement with N-body simulations (e.g., NFW). For all sets
$(a,b,d)$ in Table \ref{tbl:profiles}, the power law approximations
$C(c) \propto c^p$ and $D(c) \propto c^q$ are better than 8\% accurate
everywhere within $c = $\cmin--\cmax.

Figure \ref{fig:xi} shows $\xi(n)$ from equation (\ref{eq:xi}) for
various profiles. In all cases, $\xi(n)$ is positive, and approaches
zero as $n \rightarrow -3$. As $n \rightarrow 0$, the models all
predict a significant flattening of the $\lbol - \sigma_p$
relation. This is understandable through equation (\ref{eq:mofn}): the
characteristic density $\delta_c$ is highly anticorrelated with halo
mass in the $n \approx 0$ universes, and hence the emission measure
increases quite slowly or not at all with mass. As $n \rightarrow -3$,
$\delta_c$ becomes nearly independent of the halo mass, and therefore
$\lbol$ increases rapidly with the velocity dispersion.

\begin{deluxetable}{lccc}
\tablecaption{Density Profiles Considered}
\tablehead{\colhead{Model} & \colhead{$a,b,d$} & \colhead{$p$} &
\colhead{$q$}}
\startdata
Hernquist (1990) & 1,1,3 & 0.35   & 0.16 \nl NFW  & 1,1,2   & 0.74 & 0.58 \nl
Plummer Sphere  & 0,2,5/2 & 0.14 & 0.05 \nl King (1962) & 0,2,3/2 & 0.74 & 0.59 \nl
Jaffe (1983)   & 2,1,2 & 0.17   & 0.03 \nl
\enddata
\label{tbl:profiles}
\end{deluxetable}

\section{Application}
\label{sec:data}

For low-redshift clusters of galaxies, $Z(z,\Omega) \approx 1$ with
high accuracy. The dependence of the gas mass fraction $f$ on $\Tav$
and hence $\sigma_p$ is poorly determined. Some, using standard
analysis, claim it should increase slightly with $\Tav$ (e.g., David,
Jones, \& Forman 1995); others say that accounting for cooling flows
should cause it to decrease slightly with $\Tav$ (e.g. Allen \& Fabian
1998). Mohr, Mathiesen, \& Evrard (1999), in their detailed study
of clusters with ROSAT pointings, find that within 1 Mpc $f$ is nearly
independent of $\Tav$. Because of the range of contrasting findings, I
adopt $f \propto \Tav^{0.0 \pm 0.2} \propto \sigma_p^{0.0 \pm 0.4}$ at
the 68\% confidence level.

There is more agreement among observers regarding the empirical value
of $\alpha_3$: Quintana \& Melnick (1982) have $\alpha_3 = 4.0 \pm
0.7$ for data from the Einstein satellite; theirs are 2-10 keV
luminosities, which scale similarly to bolometric luminosities for
rich clusters. More recently, Mulchaey \& Zabludoff (1998; MZ98)
combined ROSAT observations of groups and clusters with deep optical
spectroscopy to obtain $\alpha_3 = 4.29 \pm 0.37$ for bolometric
luminosities. I adopt the average value, $\alpha_3 = 4.15 \pm 0.4$.

Assuming normally distributed errors, the adopted values of $f$ and
$\alpha_3$ constrain $\xi$ to be $< 1.0$ at the one-sided 90\%
confidence level, or $< 1.93$ at the one-sided 99\% confidence
level. Figure (\ref{fig:xi}) shows the $\xi = 1.0,1.93$
boundaries. To accommodate all the models I consider in this scenario,
$n$ must be $<-2.0$ at 90\% confidence level, and $<\nbest$ at the
99\% confidence level.

At least two systematic effects could bias these results. Preheating
of the plasma in $k \Tav \simless 4$ keV clusters (Ponman, Cannon, \&
Navarro 1999) might suppress their luminosities, while leaving those
of $k \Tav \simgreat 4$ keV clusters unchanged. One can avoid this
bias by ignoring all clusters with $k \Tav < 4$ keV. I find that
removing these clusters, which make up $\approx 25\%$ of the MZ98
sample, does not significantly affect the slope of the observed $L_X -
\sigma$ relation.

Also, cooling flows might affect the luminosities. While Markevitch
(1998) finds that the slope of the $L_X - \Tav$ relation does not
change as a result of removing cooling flows, Allen \& Fabian (1998)
find that $\alpha_2 = 3.1 \pm 0.6$ changes to $\alpha_2 = 2.3 \pm 0.4$
after including cooling flows. The Markevitch (1998) method, which
removes the cooling flow altogether, is more appropriate for probing
the gravitational potential than the Allen \& Fabian (1998) method,
which includes the luminosity of the cooling component.

To constrain the effect of cooling flows on $\alpha_3$, I conduct the
following test. Of the MZ98 clusters, 34\% are are also contained in
Markevitch (1998). I fit $\alpha_3$ for just these clusters, using the
$L_X$ which excludes the cooling component. I find that the slope
changes to $\alpha_3 = 3.85 \pm 0.3$; with this, the upper limits on
$n$ become $n<-1.7$ and $n < -0.9$ at the one-sided 90\% and 99\%
confidence levels.  The chief result of this paper---that if $n$ were
much greater than $-1$, $\alpha_3$ should be $\approx 2$ instead of
the observed value, $\approx 4$---is therefore not affected by the
10\% correction due to cooling flows.

\resizebox{3.5in}{!}{\includegraphics{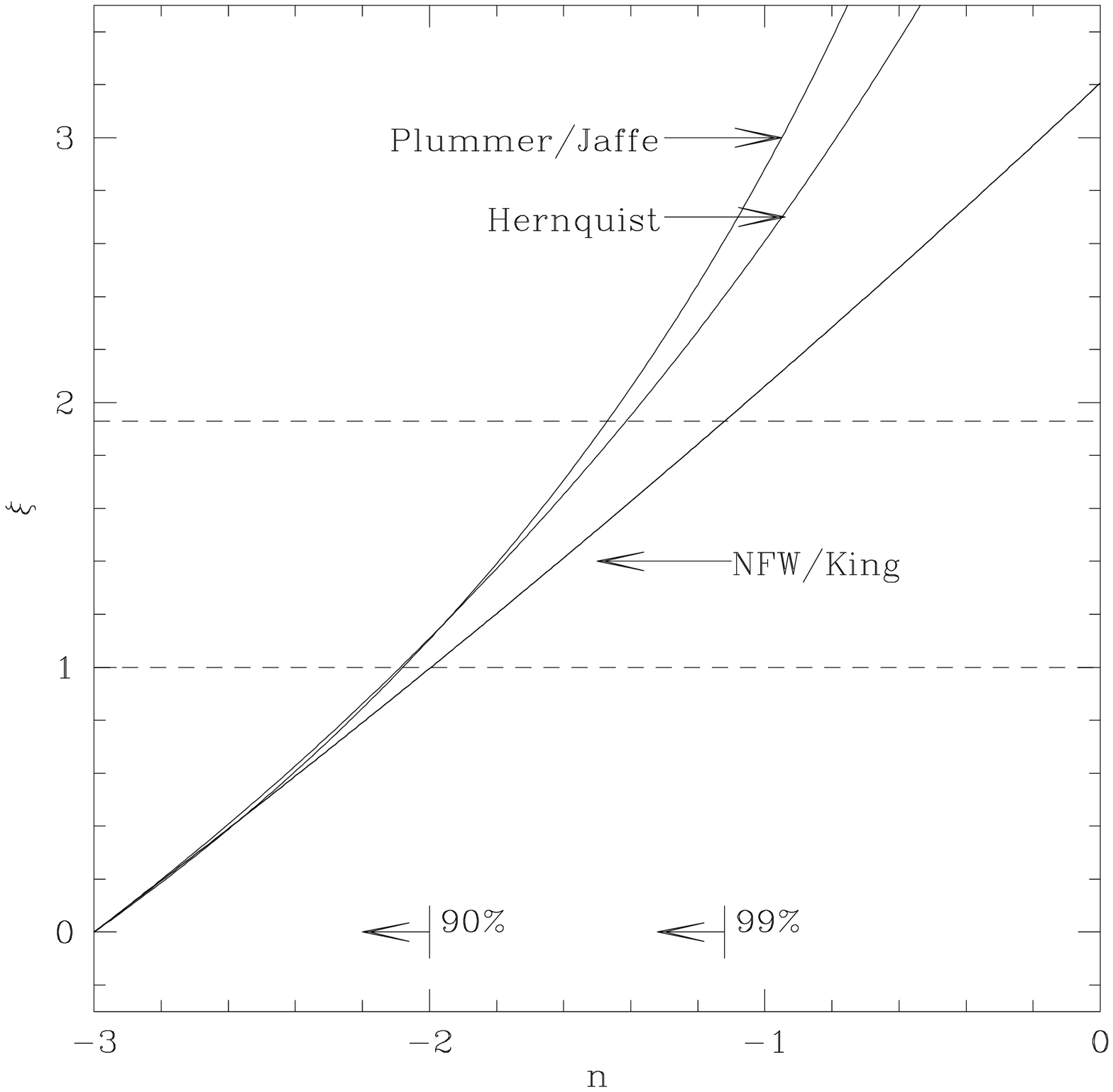}}

\figcaption[xi.eps]{The correction to the slope of the $\lbol -
\sigma_p$ relation, $\xi(n)$, for various density profiles.  The
dashed lines represent the one-sided 90\% and 99\% upper bounds on
$\xi$ from available observations. The arrows indicate the same
confidence intervals on $n$.
\label{fig:xi}}

\section{Conclusion}
\label{sec:concl}

If the characteristic densities of clusters of galaxies trace the
background density of the universe at the time of each cluster's
formation, there should be a relatively simple relationship between
the x-ray luminosity $\lbol$, the observed velocity dispersion
$\sigma_p$, and the slope of the primordial power spectrum, $n$. This
relationship, given by equations (\ref{eq:final}) and (\ref{eq:xi}),
depends slightly on the density profile of the clustered dark matter,
but should not significantly depend on $\Omega$ or $\Lambda$ when
applied to low-redshift clusters of galaxies. For a wide range of
assumed density profiles, the observations imply $n < -2.0$ and $n <
\nbest$ at the one-sided 90\% and 99\% confidence levels,
respectively. This is consistent with the bounds from the cluster
temperature function, which give $n$ between $-2.3$ and $-1.15$.
Improving this constraint depends largely on a better understanding of
preheating, cooling flows, and the variation of the gas mass fraction
with $\sigma_p$.

I thank the referee, Vincent Eke, for comments which improved the
paper. I am grateful to Margaret Geller for our many
discussions. Conversations with Saurabh Jha were also useful. This
work was supported by the Smithsonian Institution and by the National
Science Foundation.

\parindent 0in
\section*{References}

Allen, S. W., \& Fabian, A. C. 1998, MNRAS, 297, L57

Binney, J., and Tremaine, S. 1987, Galactic Dynamics (Princeton:
Princeton University Press)

Carlberg, R. G., \etal 1997, ApJ, 495, L13
 
David, L. P., Jones, C., \& Forman, W. 1995, ApJ, 445, 578

Eke, V., Navarro, J. F., \& Frenk, C. S. 1998, ApJ, 503, 569

Henry, J. P. \& Arnaud, K. A. 1991, ApJ, 372, 410

Henry, J. P. 1997, ApJ, 489, L1

Hernquist, L. 1990, ApJ, 356, 359

Jaffe, W. 1983, MNRAS, 202, 995

King, I. R. 1962, AJ, 67, 471

Mahdavi, A., Geller, M. J., B\"ohringer, H., Kurtz, M. J., \& Ramella,
M. 1997, ApJ, in press (astro-ph/9901095)

Markevitch, M. 1998, ApJ, 504, 27

Mohr, J. J., Mathiesen, N., \& Evrard, A. E. 1999, ApJ, in press
(astro-ph/9901281)

Mulchaey, J. S., \& Zabludoff, A. I. 1998, ApJ, 496, 73

Navarro, J. F., Frenk. C. S., \& White, S. D. M. 1997, ApJ, 490, 493

Ponman, T. J., Cannon, D. B., \& Navarro, J. F. 1999, Nature, 397, 135

Quintana, H., and Melnick, J., 1982, AJ, 87, 972

\end{document}